\documentclass[aps,prd,reprint,superscriptaddress,nofootinbib,floatfix,longbibliography]{revtex4-2}

\usepackage{amsmath}
\usepackage{bm}
\usepackage{amsfonts}
\usepackage{graphicx}
\usepackage{hyperref}
\usepackage{soul}
\usepackage{color}

\enlargethispage{\baselineskip}

\hypersetup{
     colorlinks   = true,
     citecolor    = blue
}
\everymath{\displaystyle}

\newcommand{\BD}{{\rm B/D}}
\newcommand{\gBL}{g_{B\textrm{--}L}}
\newcommand{\VBL}{V_{B\textrm{--}L}}
\newcommand{\f}{\mathfrak{f}}

\newcommand{\apjl}{Astrophys.~J.~Lett.}

\begin{document}

\title{Confronting the Galactic 511 keV emission with $B-L$ gauge boson dark matter}

\newcommand{\TDLI}{\affiliation{Tsung-Dao Lee Institute (TDLI) \& School of Physics and Astronomy, Shanghai Jiao Tong University, \\ Shengrong Road 520, 201210 Shanghai, P.\ R.\ China}}

\author{Weikang Lin}
\email{weikanglin@sjtu.edu.cn}
\TDLI

\author{Tsutomu T. Yanagida}
\email{tsutomu.tyanagida@sjtu.edu.cn}
\TDLI
\affiliation{Kavli IPMU (WPI), The University of Tokyo, Kashiwa, Chiba 277-8583, Japan}
\date{\today}

\begin{abstract}
The $B-L$ gauge symmetry motivated from the successful generation of the seesaw mechanism and leptogenesis. We show that if the $B-L$ gauge boson constitutes a small fraction of the dark matter (DM) it can explain the Galactic $511$ keV emission via the decay into an electron-positron pair. We find the model parameter space that is consistent with the seesaw mechanism, cosmologically viable, and accounting for the amplitude of the Galactic positron line. From this parameter space we derive an upper bound of the gauge boson mass and then an bound of the positron injection energy $\lesssim3$\,MeV. This derived energy bound is consistent with the observational upper limit of the injection energy. The resultant model predicts the $B-L$ breaking scale to be in a relatively narrow range, i.e., $V_{B\textrm{--}L}\sim10^{15}\,\textrm{--}\,10^{16}\,$GeV, which is consistent with a Grand Unification (GUT) scale seesaw mechanism. The model is consistent in several phenomenologies, suggesting their common origin from the B-L symmetry breaking. 

\end{abstract}

\maketitle

\section{Introduction} \label{sec:intro}
Since the discovery of the $511$ keV emission (presumably) from position $(e^+)$ annihilation in the Galaxy, \cite{1972ApJ...172L...1J,1975ApJ...201..593H,1978ApJ...225L..11L}, its origin has remained a mystery for about half a century; for a review, see \cite{Prantzos:2010wi}. Its large intensity and morphology are difficult to explain with astrophysical origins, which has motivated a number of DM explanations such as the light DM annihilation \cite{Boehm:2003bt,Gunion:2005rw} and decay \cite{Hooper:2004qf,Picciotto:2004rp,Takahashi:2005kp}. However, light DM candidates often suffer from other astrophysical or experimental constraints \cite{Fayet:2006sa,BaBar:2017tiz,Knapen:2017xzo,DAMIC:2019dcn}. Besides, none of these attempts are able to predict the spectral features of the Galactic $511$ keV emission line, such as the injection positron energy.

A recently proposed light DM candidate is the $B-L$ gauge boson ($A^\f_\mu$) associated with the $B-L$ gauge symmetry \cite{Choi:2020kch,Okada:2020evk,Lin:2022xbu}.\footnote{The superscript ``$\f$'' stands for ``f\'eeton''. Such a small-mass and feebly coupled gauge boson provides a unique test to the well-motivated $B-L$ gauge symmetry as is pointed out in \cite{Lin:2022xbu}. Given this important role, we have called this $B-L$ gauge boson the ``f\'eeton'' in \cite{Lin:2022xbu}.\label{ft:feeton}} The $U(1)_{B\textrm{--}L}$ gauge symmetry is a well-motivated minimum extension to the Standard Model of particle physics (SM). It is anomaly free naturally with the presence of three right-handed neutrinos. The large Majorana masses of the right-handed neutrinos are generated when such a gauge symmetry is spontaneously broken, which can simultaneously solve the problems of the small active-neutrino masses via the seesaw mechanism~\cite{Yanagida:1979as, *Yanagida:1979gs,GellMann:1980vs,Minkowski:1977sc,Wilczeck:1979CP} and the cosmological baryon asymmetry via leptogenesis \cite{Fukugita:1986hr, Buchmuller:2005eh}. When the gauge coupling constant is $\gBL\lesssim10^{-18}$, $A^\f_\mu$ can be a dark matter candidate \cite{Lin:2022xbu,Choi:2020dec}. It predominantly decays into active neutrinos, which provides a unique way to test the $B-L$ extension to SM that is otherwise difficult to probe with high-energy colliders \cite{Lin:2022xbu}. The decay to three photons is extremely suppressed \cite{Choi:2020kch,Okada:2020evk}, so that it safely satisfies the constraints from the precise $\gamma$-ray and X-ray observations. The small gauge coupling and light mass also allow it to avoid the constraints from the current DM direct searches. The next leading decay channel of $A^\f_\mu$ is that into an electron-positron pair if the  mass of $A^\f_\mu$ ($m_\f$) is larger than the threshold for such a pair production. 

Given the important role of the $B-L$ gauge boson played in testing the $B-L$ symmetry and the unresolved nature of the Galactic $511$ keV emission, it is timely to investigate whether it can explain such an astrophysical anomaly and study the implications on the physics of $B-L$ symmetry. In this work, we show that the $B-L$ gauge boson with a seesaw-motivated and cosmologically viable model parameter space can explain the amplitude of the Galactic positron line. The $A^\f_\mu$ DM can only constitute a small fraction of the total DM amount. From the resultant model, we derive an upper bound of $m_\f\simeq6$\,MeV and hence an upper bound of the positron injection energy of $\sim3$\,MeV. This derived upper bound of the positron injection energy coincides with observational the bound of positron injection energy  \cite{Beacom:2005qv}. Profoundly, the model also implies that the $B-L$ breaking scale is $\VBL=7\times10^{14}\,\textrm{--}\,10^{16}\,$GeV, which is consistent with a GUT-scale seesaw mechanism \cite{Buchmuller:1998zf,Buchmuller:2005eh} and with the energy scales required for other phenomenological considerations as we shall discuss.

\section{Positron production from the decay of $A^\f_\mu$}\label{sec:feeton_decay}
When $m_\f$ is larger than twice the electron mass $2m_e$, it decays into an electron-positron pair with a rate given by \cite{Fabbrichesi:2020wbt}
\begin{equation}\label{eq:Gamma_to_e}
    \Gamma_{\f\rightarrow e^-e^+}=\frac{\gBL^2m_{\f}}{12\pi}\sqrt{1-\frac{4m_e^2}{m_{\f}^2}}\left(1+\frac{2m_e^2}{m_{\f}^2}\right)\,,
\end{equation}
where $\gBL$ is the gauge coupling constant, $m_e$ is the electron mass. We define $\Delta m\equiv m_\f-2m_e$, and require $\Delta m>13.6$\,eV to allow positronium (Ps) formation via the charge exchange of positrons with hydrogen atoms. For the Galactic DM density distribution, we assume a Navarro–Frenk–White (NFW) DM density profile~\cite{Navarro:1995iw}, 
\begin{equation}\label{eq:NFW}
    \rho^{\textsc{g}\rm al}_{\textsc{dm}}=\frac{\rho_{\rm s}}{\frac{r}{r_{\rm s}}(1+\frac{r}{r_{\rm s}})^2}\,,
\end{equation}
with the characteristic density  $\rho_s = 1.4\times 10^7 M_\odot{\rm\,kpc^{-3}}$ and scale radius $r_s = 16\,$kpc given in \cite{Nesti:2013uwa}. We parameterize today's fraction of $A^\f_\mu$ in the DM amount as $f_\f\equiv\rho_\f^0/\rho^0_{\textsc{dm}}$, where $\rho_{\f}^0$ and $\rho^0_{\textsc{dm}}$ are the cosmic average density of the $A^\f_\mu$ DM and the total DM today. We assume this fraction holds universally for the entire universe. The positron production rate is then $\dot{n}_{e^+}=\Gamma_{\f\rightarrow e^-e^+}f_\f\rho^{\textsc{g}\rm al}_{\textsc{dm}}/m_\f$. 

\begin{figure}[tbp]
    \centering
    \includegraphics[width=\linewidth]{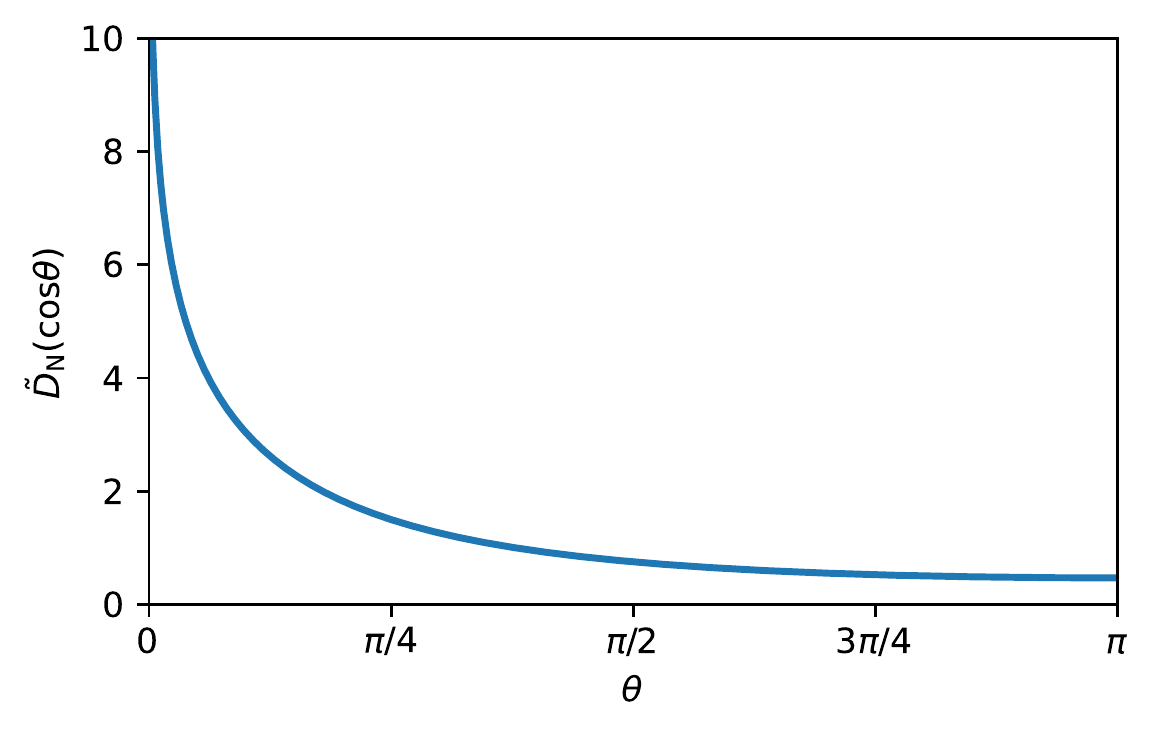}
    \caption{The angular dependence of function $\tilde{D}_{\rm N}(\cos\theta)$ defined in Eq.\,\eqref{eq:positron-annihilation}.}
    \label{fig:D_function}
\end{figure}

After the positrons are produced, they can annihilate with electrons either directly or via Ps formations. The Ps fraction $f_{Ps}$---the ratio of the number of positron forming positronium to the total number---is measured to be close to unity \cite{Harris:1998tt,Siegert:2015knp,Jean:2005af}. Here, we take $f_{Ps}=1$ for simplicity.\footnote{We keep in mind that a somewhat lower $f_{Ps}=0.76 \pm 0.12$ is reported recently in \cite{Kierans:2019aqz}. Adopting this different $f_\f$ does not qualitatively change our conclusions.} For all the Ps, 1/4 of them are in the  para-positronium (p-Ps) state that annihilates into two photons with $511$ keV. We assume the positrons annihilate closely to their production sites and we equal the positron annihilation rate to the production rate. The production rate of the $511$ keV photons is then $\dot{n}_{\gamma}=2\dot{n}_{e^+}/4$, and the angular differential flux is given by the following integral along the line of sight ($s$),
\begin{align}
    \frac{{\rm d}\Phi_{511}}{{\rm d}\Omega}&=\frac{1}{4\pi}\int\dot{n}_{\gamma} {\rm d}s\nonumber\\
    &=4\times10^3f_\f\left(\frac{\gBL}{10^{-20}}\right)^2\sqrt{1-\frac{4m_e^2}{m_{\f}^2}}\left(1+\frac{2m_e^2}{m_{\f}^2}\right) \nonumber\\
    &~~~~~~~~~~~~\times  \tilde{D}_{\rm N}(\cos\theta)\,[{\rm cm}^{-2}{\rm s}^{-1}{\rm sr}^{-1}]\,,\label{eq:positron-annihilation}
\end{align}
where $\tilde{D}_{\rm N}(\cos\theta)$ is a function of the angle ($\theta$) from the Galactic Center (GC) representing the morphology of the flux and is normalized so that $\int\tilde{D}_{\rm N}(\cos\theta)\,{{\rm d}\Omega}=4\pi$ with $\cos\theta = \cos\ell\cos b$ where $\ell$ and $b$ are Galactic longitude and latitude. The function $\tilde{D}_{\rm N}(\cos\theta)$ is plotted in Figure \ref{fig:D_function}. There are then three model parameters ($f_\f$, $\gBL$, $m_\f$). 

\section{Comparison to the Galactic 511 keV emission}\label{sec:comparison_to_data}
The Galactic $511$ keV emission has a rather diffuse morphology but is more concentrated towards GC than radiation in other wavelengths \cite{2010ApJ...720.1772B,Siegert:2015knp}. Analyses of the associated Bremsstrahlung and in-flight annihilation radiations indicate that the injection energy of the positron is  $\lesssim3$ MeV \cite{Beacom:2005qv,Sizun:2006uh}. Table \ref{tab:Measurements} summarizes some important parameters about the Galactic $511$ keV emission given in \cite{Siegert:2015knp} and \cite{Beacom:2005qv}.

\begin{table}[tbp]
\caption{Parameters of the Galactic $511$ keV emission adopted from \cite{Siegert:2015knp}. The last row is from \cite{Beacom:2005qv}. }
\label{tab:Measurements}
\begin{ruledtabular}
\begin{tabular}{ll}
Field & Value \\
\hline
Total intensity & $2.74\pm0.25\times10^{-3}$\,cm$^{-1}$s$^{-1}$ \\
Bulge intensity & $0.96\pm0.07\times10^{-3}$\,cm$^{-1}$s$^{-1}$\\
Disk intensity & $1.66\pm0.35\times10^{-3}$\,cm$^{-1}$s$^{-1}$\\
Bulge/disk ratio & $0.58\pm0.13$\\
Bulge extent $(\sigma_\ell,~\sigma_b)$& $(8.7,\,8.7)$ [degrees]\\
Disk extent $(\sigma_\ell,~\sigma_b)$ &
$(60^{+10}_{-5},10.5^{+2.5}_{-1.5})$ [degrees]\\
Ps fraction $f_{Ps}$ (bulge) & $1.080\pm0.029$\\
Injection energy of $e^+$ & $\lesssim3$\,MeV 
\end{tabular}
\end{ruledtabular}
\end{table}

\begin{figure*}[htbp]
    \centering
    \includegraphics[width=0.99\textwidth]{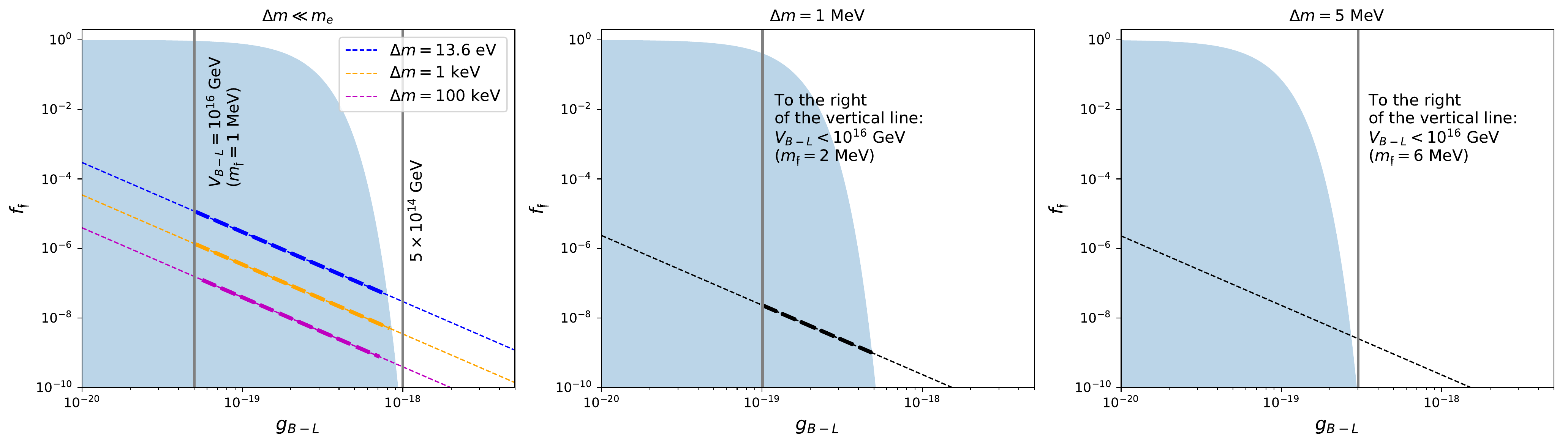}
    \caption{The parameter space in cases of different $m_\f$ (and $\Delta m\equiv m_{\f}-2m_e$). Left: $\Delta m\ll m_e$. The light blue region is the cosmologically viable parameter space. The vertical line indicates the value of $\gBL=m_\f/(2\VBL)$ for some specific values of $\VBL$. The dashed lines are parameters that can account for the bulge positron annihilation flux, while their thick portions, in addition, satisfy the cosmological constraints and the motivation from the seesaw mechanism. Middle: $m_\f=2$\,MeV. Right: $m_\f=6$\,MeV. There is no viable parameter space when $m_\f\gtrsim6$\,MeV.  }
    \label{fig:ParamSpace}
\end{figure*}

\subsection{The bulge flux and constraints on model parameters}\label{sec:bulge_flux}
We first focus on the bulge flux and will discuss the flux morphology later. In \cite{Siegert:2015knp}, the (broad) bulge region is modeled as a two-dimensional Gaussian function with a width $\sigma_\ell=\sigma_b=8.7$\,degrees, or a Full-Width-at-Half-Magnitude (FWHM) of $20.55$\,degrees in either dimension.\footnote{We are aware of that the bulge region is modeled with a narrow bulge and a broad bulge \cite{Siegert:2015knp}. To calculate the bulge flux, we only refer to the broad bulge region for the bulge area and we compare the predicted bulge flux to the total measured bulge intensity.} Using Eq.\,\eqref{eq:positron-annihilation}, we calculate the integrated flux for a region that is within $\theta<10.28$ degrees from GC and equal it to half of the measured bulge flux. With $\int_{\theta<10.28^\circ}\tilde{D}_{\rm N}(\cos\theta)\,{{\rm d}\Omega}=0.52$ and the bulge intensity given in Table \ref{tab:Measurements}, the above procedure gives us the following constraint on the model parameters,
\begin{equation}\label{eq:bulge_flux_constraint}
    f_\f\left(\frac{\gBL}{10^{-20}}\right)^2\sqrt{1-\frac{4m_e^2}{m_{\f}^2}}\left(1+\frac{2m_e^2}{m_{\f}^2}\right)\simeq2.3\times10^{-7}\,.
\end{equation}
In Figure \ref{fig:ParamSpace}, the dashed lines show this constraint in the ($f_\f$,\,$\gBL$) plane for different $\Delta m$'s. The value of $f_\f$ for a fixed $\gBL$ is smaller as $\Delta m$ (or equivalently, $m_\f$) becomes larger, but it becomes insensitive to $\Delta m$ when $\Delta m\gtrsim m_e$. This is because, when $\Delta m\ll m_e$, the predicted flux increases rapidly with $\Delta m$ [see Eq.\,\eqref{eq:positron-annihilation}] and $f_\f$ needs to decrease to match the observed value of the flux. On the other hand, when $\Delta m\gtrsim m_e$, the predicted flux becomes insensitive to $m_\f$. We find that $f_\f$ is typically very small, so $A^\f_\mu$ only constitutes a small fraction of the DM amount. 

Secondly, we require the $A^\f_\mu$ DM to satisfy cosmological observations. Given the small fraction of $A^\f_\mu$ in the DM amount today, to estimate such a cosmological constraint we require that $A^\f_\mu$ has never been the dominant DM component in the past, i.e., the comoving density $\rho_\f(t)=\rho_\f^0\exp[(t_{\textsc{u}}-t)/\tau]<\rho^0_{\textsc{cdm}}$ at all times\footnote{The $A^\f_\mu$ DM can be the dominant DM component well before the epoch of matter-radiation equality. But substituting such a small initial time instead of $t=0$ does not affect our estimation.} ($t$), where $\tau$ is the lifetime of $A^\f_\mu$, $t_{\textsc{u}}$ is the age of the universe and ``CDM'' denotes the dominant DM component. Since $\rho_\f^0\ll\rho^0_{\textsc{cdm}}$, the estimated cosmological constraint translates into
\begin{equation}\label{eq:cosmological-constraint}
    f_\f\exp(t_{\textsc{u}}/\tau)<1\,.
\end{equation}
We note that the above estimate of the cosmological constraint is rather conservative. Given the consistency between the DM density measured at the early times and that measured at the late times \cite{Lin:2021sfs,Planck:2018vyg,Pan-STARRS1:2017jku}, we expect the detailed cosmological constraint is stronger than our estimate. But, our estimate suffices for the goal of this work. This is because, as we shall see, the allowed $f_\f$ value drops very quickly as $\gBL$ increases and it is the range of $\gBL$ that is relevant for the inference of $\VBL$.

Recall that $A^\f_\mu$ predominantly decays into the three active neutrinos with a rate given by \cite{Lin:2022xbu}
\begin{equation}
    \Gamma_{\f\rightarrow \nu\bar{\nu}}=\frac{\gBL^2}{8\pi}m_{\f}\,.
\end{equation}
The lifetime of $A^\f_\mu$ is then $\frac{1}{\tau}=\Gamma_{\f\rightarrow \nu\bar{\nu}}+\Gamma_{\f\rightarrow e^-e^+}$. The light blue regions in Figure \ref{fig:ParamSpace} show the cosmologically viable parameter space for cases with different $m_\f$'s. This region shrinks as $m_\f$ increases, which can be seen from the left to the right panels.

Finally, the $B$-$L$ breaking scale $\VBL$ is of the order of $\VBL=10^{12}\,\textrm{--}\,10^{16}\,$GeV for the successful generation of the seesaw mechanism and the leptogenesis. Since $m_\f=2\gBL\VBL$, the range of $\gBL$ corresponding to $\VBL=10^{12}\,\textrm{--}\,10^{16}\,$GeV changes with $m_\f$. For each panel in Figure \ref{fig:ParamSpace} we show the $\gBL$ value corresponding to $\VBL=10^{16}\,$GeV with a vertical line and the theoretically motivated parameter space is that to the right. 

Combing the above three constraints, the viable parameters are shown by the thick portions of the dashed lines in Figure \ref{fig:ParamSpace}. In the left panel, we show some examples for $\Delta m\ll m_e$. In this case, as already discussed, $f_\f$ decreases as $m_\f$ increases. On the other hand, the range of $\gBL$ that is cosmologically viable (light blue) and corresponds to $\VBL<10^{16}\,$GeV (right to the vertical line) is only slightly affected.\footnote{In the left panel, we only show the $m_\f=1$ MeV case for the light blue region and the vertical line, but the effect of $\Delta m$ can be seen by the slightly shrinking thick dashed lines as $\Delta m$ increases.} The fraction of $A^\f_\mu$ in DM is found to be $f_\f\sim10^{-9}\,\textrm{--}\,10^{-5}$.

In the middle and the right panels of Figure \ref{fig:ParamSpace}, we show two examples of $\Delta m\gtrsim m_e$ where the effects from an increasing $m_\f$ are opposite to the case when $\Delta m\ll m_e$. The value of $f_\f$ is now insensitive to $m_\f$ because the predicted flux becomes insensitive to $m_\f$. So, the dashed lines in these two panels are essentially the same. However, the range of $\gBL$ shrinks as $m_\f$ increases. As a consequence, there is no more viable parameter space for $m_\f\gtrsim6$\,MeV; see the right panel. 

Interestingly, releasing $f_\f$ as a free parameter does not drastically changes the upper limit of $m_\f$ compared to that found in \cite{Lin:2022xbu}. In fact, the bound on $m_\f$ is very robust against the value of $f_\f$ as long as $A^\f_\mu$ contributes some amount of DM today. For example, even if $f_\f$ is as low as $10^{-20}$, the upper limit of $m_\f$ is only slightly changed to $m_\f\lesssim7$\,MeV. 

\subsection{The upper limit of the positron injection energy and the $B-L$ breaking scale}
From the above analyses, we derive two important implications if the $B-L$ gauge boson DM is the source of the Galactic $511$ keV emission anomaly.

1. Given the upper limit of $m_\f\lesssim6$\,MeV, the model predicts the positron injection energy to be $\lesssim3$\,MeV. This is consistent with observations  \cite{Beacom:2005qv,Sizun:2006uh}. We note that none of the other current DM scenarios are able to predict this feature of the Galactic positron annihilation emission, which is also difficult to reproduce with some (but not all) astrophysical sources \cite{Prantzos:2010wi}.

2. The inferred $B-L$ breaking scale is rather narrow, i.e., $\VBL=7\times10^{14}\,\textrm{--}\,10^{16}\,$GeV, which can be read from the left panel of Figure \ref{fig:ParamSpace}. This is much improved and more informative than the naive estimation from the seesaw mechanism alone ( $10^{12}\,\textrm{--}\,10^{16}\,$GeV). The inferred range of $\VBL$ is consistent with a GUT-scale seesaw \cite{Buchmuller:1998zf,Buchmuller:2005eh}. We further remark on the significance of such a range of $\VBL$ below. 

Such a high energy scale of $\VBL$ is difficult to reach with the current and near-future particle colliders. It is then important to investigate tests with other indirect astrophysical phenomenology. It is pointed out in \cite{Lin:2022xbu} that the detection of neutrinos from decays of $A^\f_\mu$ would be a smoking gun for the $B-L$ symmetry extension to SM. For the scenario considered in this work, however, the fraction of $A^\f_\mu$ in DM is so small that it is not realistic in the near future to detect such a neutrino signal.

Alternatively, one can study the $A^\f_\mu$ DM production mechanism and the early-universe phenomenology. Remarkably, besides being consistent with a GUT-scale seesaw, the inferred range of $\VBL$ is consistent with some energy scales under other phenomenological considerations in the early universe. 

First, if the $B-L$ symmetry is broken during inflation, the $B-L$ gauge boson DM as a vector boson can be produced during inflation without violating the constraint from the non-detection of isocurvature perturbations \cite{Graham:2015rva}. The viable parameter space here permits a self consistency for such an inflationary production: if produced during inflation,  $f_\f$ is related to $m_\f$ and the inflation scale $H_{\rm inf}$ by $f_\f=\big(\frac{m_\f}{6\times10^{-9}\,{\rm keV}}\big)^{1/2}\big(\frac{H_{\rm inf}}{10^{14}\,{\rm GeV}}\big)^2$ \cite{Graham:2015rva}. Taking $f_\f\lesssim10^{-5}$ and $m_\f\simeq1$\,MeV, we obtain an inflation scale of $H_{\rm inf}\lesssim5\times10^8$\,GeV. This is in turn consistent with the condition that the $B-L$ symmetry is broken during inflation, because $H_{\rm inf}<\VBL$. In addition, the string axion DM can be accommodated as the major DM component, since the inferred inflation scale satisfies the constraint from the non-detection of the isocurvature perturbation; see Eq.\,(6) in \cite{Kawasaki:2015pva}. 

Further, on top of the above inflationary production, if the quartic coefficient ($\lambda$) of the scalar field ($\phi$) responsible for the spontaneous symmetry breaking is $\lambda\lesssim10^{-4}$, the reheating temperature can be higher than the mass of the scalar field. In that case, the $B-L$ symmetry may restore after reheating and be broken again as the temperature cools down. Cosmic string loops can form due to such a phase transition after reheating \cite{Kibble:1976sj}. Those cosmic strings emit GWs as they shrink and loose energy \cite{PhysRevD.31.3052}, which may explain the recently reported detection of a stochastic GW background by NANOGrav \cite{Blasi:2020mfx,Ellis:2020ena}. Interestingly, the inferred range of $\VBL$ required to source the Galactic $511$ keV emission coincides with that required to explain the NANOGrav detection; see Eq.\,(10) in \cite{Blasi:2020mfx}. This scenario can be further tested with future GW experiments such as SKA \cite{Janssen:2014dka} and LISA \cite{Bartolo:2016ami}. We leave the full exploration of the early-universe phenomenology in a future work.

\subsection{Remarks on the flux morphology}\label{sec:morphology}
The morphology of the Galactic $511$ keV emission has been difficult to explain with all astrophysical or DM sources \cite{Prantzos:2010wi}. Such an emission is concentrated towards GC, which can be roughly presented by a high value of bulge-to-disk flux ratio $\BD$. Based on the earlier data from INTEGRAL/SPI with $\BD\sim1.5$ \cite{Knodlseder:2005yq}, it was found that decaying DM scenarios are disfavored unless the inner DM density increases towards the center very sharply \cite{Ascaisbar:20006apil}. A similar conclusion was obtained in \cite{Skinner:20156m} where it was found that the positron production rate is proportional to the DM density squared. This morphology problem to decaying dark matter scenarios (and to the traditional astrophysical explanations in general) is alleviated with the new data as the $\BD$ has reduced to $0.58\pm0.13$ \cite{Siegert:2015knp}. The flux morphology in the decaying DM scenarios is solely described by the function $\tilde{D}_{\rm N}(\cos\theta)$. We estimate the $\BD$ for decaying DM scenarios by taking (the mean values of) the extents of the bulge and the disk derived in \cite{Siegert:2015knp}, which is summarized in Table \ref{tab:Measurements}. We obtain $\BD\simeq0.3$, which is still about a factor of $2$ (and $\sim2$-$\sigma$) smaller than the derived value from observations \cite{Siegert:2015knp}.\footnote{For a more robust conclusion one should compare the predicted flux profile with the observed flux profile, which is beyond the goal of this work.} Therefore, decaying DM scenarios are still in mild tension with the new data. 

It is however too early to exclude decaying DM scenarios based on the currently measured morphology of the flux for two reasons. (1) Given some uncertainties of the positron transportation in the interstellar medium \cite{Prantzos:2010wi}, the assumption that positrons annihilate closely to their production sites may not be satisfied in the disk area. Some positrons may have escaped from the disk, reducing the annihilation flux from there and giving a larger predicted $\BD$. (2) Due to the low surface luminosity of the flux from the disk, its detection has proven to be difficult \cite{Skinner:20156m,Siegert:2015knp,Kierans:2019aqz}. It is possible that the detection of the flux from the disk is still incomplete and the actual disk flux is larger, and hence the current observed $\BD$ might be biased to be larger than the true value. Thus, it is still possible that DM decays can explain the morphology of the Galactic $511$ emission, which we assume in this work. Since the detected bulge flux is more reliable, we only use it to infer the model parameters as we did in the previous sections.


\section{Conclusion}
The unresolved nature of the Galactic $511$ keV emission could point to new physics beyond SM. In this work, we have explored a scenario that the decay of the $B-L$ gauge boson DM into electron-positron pairs sources such an emission. We consistently consider the model parameter space that is theoretically motivated from the seesaw mechanism, viable with cosmology and accounting for the Galactic $511$ keV bulge emission. We find that the resultant model successfully accounts for the positron injection energy and has an important prediction on the physics of the $B-L$ symmetry.

We find that, while the fraction of $A^\f_\mu$ in the DM amount is released as a free parameter, its mass is bounded to $\lesssim6$\,MeV by the cosmological constraint and the scale of the $B-L$ symmetry breaking. This bound is very robust against the value of $f_\f$. The $A^\f_\mu$ DM is found to constitute only a small fraction of the DM amount ($f_\f\sim10^{-9}\,\textrm{--}\,10^{-5}$).

As a result of the constraint on $m_\f$, the positron injection energy from the decay of $A^\f_\mu$ is bounded to $\lesssim3$\,MeV, which coincides with the current observational limit of the injection energy. This is different from other DM scenarios in the literature so far: In our case, we derive an upper bound of $m_\f$ and hence an upper bound of the positron injection energy from the consideration of the model consistency with cosmology, the seesaw mechanism and the amplitude of the Galactic positron line. This derived energy bound turns out to be consistent with the observational limit of the positron injection energy. On the contrary, other DM scenarios need to use the upper bound of the positron injection energy to set a constraint on the DM mass range.


The model has a nontrivial implication on the $B-L$ physics: the $B-L$ symmetry breaking scale is predicted to be $\VBL=7\times10^{14}\,\textrm{--}\,10^{16}\,$GeV, which is consistent with the GUT-scale seesaw mechanism. The range of $\VBL$ permits a self-consistent inflationary production of the $A^\f_\mu$ DM and accommodates the possibility that the cosmic strings generated by the gauge $U(1)_{B\textrm{--}L}$ breaking after reheating explain the stochastic GW background reported by NANOGrav. The resultant model is consistent in several phenomenologies including the small neutrino masses, the cosmic baryon asymmetry, the Galactic $511$ keV emission and the tentative stochastic GW background reported by NANOGrav, which suggests their common origin from the B-L symmetry breaking.

One caveat is that we assume the morphology of the Galactic $511$ keV can be explained by DM decays. There is still some mild tension between the flux bulge-to-disk ratio predicted in decaying DM scenarios and that derived from current observations. However, we argue that the tension may be alleviated or even eliminated with further studies of the transportation of positrons in the interstellar medium and more complete surveys in the disk area. 

We note that our definition of $U(1)_{B\textrm{--}L}$ is not to be confused with the generalization that includes the SM hypercharge \cite{Okada:2020evk}. However, as long as the decay of $A^\f_\mu$ into neutrinos is not suppressed, our conclusions are not significantly changed.

Finally, we comment on the small gauge coupling constant which is another caveat of our work. However, while the B-L gauge symmetry is well motivated by the seesaw mechanism and leptogenesis, the gauge coupling constant is completely undetermined by theory and needs to be determined by experiments. Here, $\gBL$ is bounded by assuming the $B-L$ gauge boson is a long-lived DM component. Second, the small gauge coupling is not a result of an extreme fine-tuning. Since $U(1)_{B\textrm{--}L}$ is an asymptotic non-free theory, $\gBL=0$ is the infrared stable point and thus, although some tuning is needed to fit observations (just like any other parameter in a model), a small $\gBL$ is natural. This may be different from $U(1)_Y$, since the hypercharge $U(1)_Y$ might be embedded into the GUT group $SU(5)$. In that case, $U(1)_Y$ is a part of the asymptotic free gauge theory.

\begin{acknowledgments}
 T.\ T.\ Y.\ is supported in part by the China Grant for Talent Scientific Start-Up Project and by Natural Science Foundation of China (NSFC) under grant No.\ 12175134 as well as by World Premier International Research Center Initiative (WPI Initiative), MEXT, Japan. 
\end{acknowledgments}

\bibliographystyle{apsrev4-1}
\bibliography{refs}

\end{document}